\documentclass[aps,prd,twocolumn,showpacs,preprintnumbers,superscriptaddress]{revtex4-1} 
\usepackage{dcolumn} 
\usepackage{graphicx}
\usepackage{slashed,color}
\usepackage{amsmath,amsfonts,bm}
\usepackage{amssymb}
\usepackage{calligra}
\usepackage[T1]{fontenc}

\newcommand{\nn}{\nonumber \\ }

\usepackage{yfonts}
\newcommand{\beq}{\begin{equation}}
\newcommand{\eeq}{\end{equation}}
\newcommand{\bea}{\begin{eqnarray}}
\newcommand{\eea}{\end{eqnarray}}

\def\OMIT#1{{}}

\begin{document}


\title{One-loop evolution of parton pseudo-distribution functions on the   lattice}

 \author{Anatoly Radyushkin }
\affiliation{Physics Department, Old Dominion University, Norfolk,
             VA 23529, USA}
       \affiliation{Thomas Jefferson National Accelerator Facility,
              Newport News, VA 23606, USA}

\begin{abstract}

  \noindent
  
  We  incorporate recent  calculations of   one-loop corrections 
  for the reduced Ioffe-time pseudo-distribution ${\mathfrak M} (\nu,z_3^2)$  to 
  extend    the leading-logarithm 
   analysis of lattice data obtained by Orginos et al.  
We   observe  that the one-loop   corrections contain a  large term
reflecting the fact that effective distances involved in  the  most important diagrams
    are much smaller than the nominal distance $z_3$. 
    The 
     large  correction in this  case may be 
    absorbed into the  evolution term, and 
    the perturbative expansion used for extraction of parton densities  
    at the $\mu \approx 2$ GeV scale 
      is  under control.
    The   extracted   parton distribution   is rather close to 
      global fits  in the $x>0.1$ region,  but deviates from them for  $x<0.1$.

\end{abstract}



\maketitle


\section{ Introduction}

Feynman's parton distribution functions (PDFs) \cite{Feynman:1973xc}  $f(x)$ 
are  the crucial building blocks in the description of hard inclusive processes 
in quantum chromodynamics (QCD).
Accumulating nonperturbative  information about 
the hadron structure,   the PDFs  are a natural subject for 
a lattice study. 
 However, straightforward definitions of  PDFs 
refer to    matrix elements
of bilocal operators on   the light cone $z^2=0$,
the intervals inaccessible on   the   Euclidean lattice.

The ideas of how to    get  information  from space-like intervals 
date to the pioneering paper of  \mbox{W. Detmold}   and D. Lin 
\cite{Detmold:2005gg}  who proposed a   lattice study 
of  the deep-inelastic-type Euclidean correlators   of heavy-light  currents.
Later, V. Braun and D. M\"uller 
\cite{Braun:2007wv}
proposed to use   Euclidean correlators to
extract  the pion 
distribution amplitude,
another   function \cite{Radyushkin:1977gp} playing a fundamental role in  perturbative QCD studies 
of  hard exclusive processes. 
The use of correlators in the form of ``lattice cross sections''  was more recently 
  advocated in the papers
by Qiu and collaborators \cite{Ma:2014jla,Ma:2017pxb}.

The current correlators involve a quark propagator connecting the current vertices.
This factor is avoided in   the  proposal   by X. Ji   \cite{Ji:2013dva}  to study  the 
quasi-PDFs $Q(y,p_3)$  that describe the  distribution
of the spatial $z_3$-component of the hadron momentum $p_3$.
While being different from the Feynman PDFs $f(y)$ describing the distribution
of the hadron's ``plus''-momentum $p_+=p_0+p_3$,
they  coincide with  $f(y)$ in  the infinite momentum limit
$p_3\to \infty$. 

Both on the lattice and in the usual continuum space,
the basic object for all types of PDFs is the 
matrix element $M(z,p)$ generically 
(i.e. ignoring the inessential spin complications)
 written as $\langle  p | \phi (0)  \phi(z)  | p \rangle $.
 By Lorentz invariance, it is a function of the 
 Ioffe time $(pz)\equiv -\nu$ \cite{Ioffe:1969kf} and the interval $z^2$,
 $M(z,p) \equiv {\cal M} (\nu, -z^2)$.
 
 In the (formal)  light-cone  limit $z^2=0$, the Fourier transform
 of  $ {\cal M} (\nu, 0)$ with respect to $\nu$ 
 gives $f(x)$.  In this sense, the $\nu$-dependence of 
 the Ioffe-time distribution (ITD) ${\cal M} (\nu, -z^2)$ reflects 
 the longitudinal structure of the PDFs.
 As shown in \mbox{Ref. \cite{Radyushkin:2016hsy}}, 
 the \mbox{$z^2$-dependence}  of  ${\cal M} (\nu, -z^2)$ determines
 the \mbox{$k_\perp$-dependence}  of the  (straight-link in the case of QCD) 
  transverse momentum dependent
 parton distributions (TMDs) ${\cal F} (x, k_\perp)$.

 Since the quasi-PDFs $Q(y,p_3)$ are given by
 the Fourier transform of ${\cal M} (z_3 p_3, z_3^2)$
 with respect to $z_3$, their shape  is distorted by nonperturbative  
 transverse momentum effects  entering through the second argument of 
  ${\cal M} (z_3 p_3, z_3^2)$.  
 While,   in a general perspective, the 
 \mbox{$k_\perp$-dependence}  of  ${\cal F} (x, k_\perp)$
provides information about the three-dimensional structure 
of hadrons, in the case of the quasi-PDFs it 
is a nuisance responsible for the unwanted  difference between
$Q(y,p_3)$ and $f(y)$ that is very strong at momenta reached in 
existing lattice calculations of quasi-PDFs. 

To decrease the impact of the $z^2$-dependence of the 
ITD ${\cal M} (\nu, -z^2)$, it was proposed  \cite{Radyushkin:2017cyf}
to consider the reduced ITD ${\mathfrak M} (\nu, -z^2)$
given by the ratio of  ${\cal M} (\nu, -z^2)$ and the rest-frame 
distribution  ${\cal M} (0, -z^2)$. 
Though  there are no first-principle grounds
that the nonperturbative part of the \mbox{$z^2$-dependence}  disappears in this ratio,
it is natural to expect that it is strongly reduced.

The ideal case when ${\mathfrak M} (\nu, -z^2)$ is  just a function 
of $\nu$  corresponds to factorization
of the $x$ and $k_\perp$ dependencies of the TMD 
${\cal F} (x, k_\perp)$.  In fact, the idea that ${\cal F} (x, k_\perp) = f(x) K(k_\perp)$
in the soft region $k_\perp^2 \lesssim 1$ GeV$^2$  is 
a standard assumption of the TMD practitioners  (see, e.g., Ref. \cite{Anselmino:2013lza}), 
with a Gaussian being the most popular form for 
$K(k_\perp)$.  

An exploratory lattice study of the reduced ITD 
was  performed in Ref. \cite{Orginos:2017kos} (and also described in Ref.
\cite{Karpie:2017bzm}).  The results show that
${\mathfrak M} (\nu, z_3^2)$ is basically a universal 
function of $\nu$, with small deviations from the common curve 
for the points corresponding to the smallest values of $z_3$. 

As demonstrated in Ref. \cite{Orginos:2017kos}, these 
deviations may be explained by  perturbative evolution. 
While  the leading logarithmic approximation (LLA) 
used in Ref.  \cite{Orginos:2017kos}  is  sufficient to analyze the
$\ln z_3^2$ dependence, 
one needs to go   beyond it to
 specify the scale $\mu$  which should be attributed 
to the extracted scale-dependent PDFs $f(x,\mu^2)$.

To this end,  one needs complete expressions for one-loop corrections to ITDs.
Recently,  such  calculations have been reported  in Refs. 
 \cite{Ji:2017rah,Radyushkin:2017lvu}.   
Our goal here is to give a more detailed discussion 
of the LLA  treatment of the evolution,  
and also to extend the  analysis 
beyond the LLA.  
As we will  show,  the one-loop correction contains a large
 contribution  that  considerably changes the results 
 obtained in the LLA.

 To make this article  self-contained, we outline  in \mbox{Sec. II} 
 the basics of  the Ioffe-time distributions and pseudo-PDFs. 
 In Sec.  III, we discuss the structure of  one-loop corrections. 
 In Sec.  IV, we  describe the evolution effects 
 revealed  in the lattice study of Ref. \cite{Orginos:2017kos},
 and convert the data for  the reduced ITD ${\mathfrak M} (\nu, z_3^2)$
 into the standard parton densities $f(x,\mu^2)$ defined in the $\overline{\rm MS}$  scheme. 
 The summary of the paper and conclusions  are  given in \mbox{Sec.  V.}

 \section{Ioffe-time  distributions and pseudo-PDFs}

 The basic object for defining parton distributions is a matrix element 
 of a bilocal operator that  (skipping inessential details of  its 
 spin structure)   may be written   generically  like 
 $\langle  p | \phi (0)  \phi(z)  | p \rangle $. 
Due to invariance under Lorentz transformations, it is given by a function of two scalars,  
the  {\it Ioffe time}   $(pz)$  \cite{Ioffe:1969kf}  
(which will  be denoted by $-\nu$)  and  the interval  $z^2$
  \begin{align}
  \langle p |   \phi(0) \phi (z)|p \rangle 
=  & {\cal M} (-(pz), -z^2)  =  {\cal M} (\nu, -z^2)  
\,  
 \
 \label{lorentz}
\end{align} 
 (again, the sign for the second argument   is   chosen  so as to have  
  a positive value  for \mbox{spacelike $z$)}.
 One can demonstrate \cite{Radyushkin:2016hsy,Radyushkin:1983wh}    that, for all relevant Feynman diagrams,   
 its  Fourier transform  ${\cal P} (x, -z^2)$ with respect to $(pz)$ 
 has $-1 \leq x \leq 1$ as support, i.e., 
   \begin{align}
 {\cal M} (-(pz), -z^2) 
&   = 
 \int_{-1}^1 dx 
 \, e^{-i x (pz) } \,  {\cal P} (x, -z^2)  \   .
  \label{MPD}
\end{align}   
   In this  covariant  definition of  $x$,  one does not need to assume that  
   $z$ is on the light cone
$z^2=0$ or that $p$   is light-like $p^2=0$.
  
  On the light cone $z^2=0$, we formally have 
 \mbox{$   {\cal P} (x, 0) =f(x) $.}  Hence, the function $ {\cal P} (x, -z^2) $ 
 may be treated as   a generalization of the concept of 
PDFs onto non-lightlike intervals $z^2$, and  following   \cite{Radyushkin:2017cyf} ,  we will  refer to it as the
 {\it pseudo-PDF}.  In view of lattice applications, we will take the separation 
$z=\{0,0,0,z_3\}$ oriented  in the direction specified by the hadron momentum $p=\{E,0,0,P\}$.

  In renormalizable theories (including QCD),    
  the function  $ {\cal M} (\nu, -z^2)  $ 
 has  logarithmic $\sim \ln (-z^2)$  singularities.
 In deep inelastic scattering (DIS), they  result
 in a logarithmic  scaling violation with respect 
 to the photon virtuality  $Q^2$. 
 A wide-spread  statement is that the \mbox{$Q^2$-dependent}   DIS structure functions 
  $W(x_B,Q^2)$  probe the   hadron structure 
 at distances $\sim 1/Q$.  
 In the case  of the pseudo-PDFs ${\cal P} (x,z_3^2)$, 
 one may say that  they literally describe the hadron structure 
 at  the distance $z_3$.
 
 Just like the  DIS  form factors  $W(x_B,Q^2)$ are  
 written in terms of  the universal parton densities $f(x, Q^2)$,  
 the pseudo-PDFs   obtained   from 
 lattice calculations   may  be expressed through  the usual 
 parton distributions. The  latter are  defined by 
 the operators on the light cone $z^2=0$, i.e., in a 
 logarithmically  singular limit.
  In the approach based on  the operator product expansion
(OPE),  the standard procedure is to  remove  these singularities  with the help of some 
 prescription. 
 
 The most popular of them is the  $\overline{\rm MS}$  scheme based 
on the dimensional  regularization.  
Consequently,  the resulting PDFs  have a dependence on the renormalization scale 
$\mu$, and therefore one should write the PDFs as $   f(x, \mu^2)$.  
Switching from $x$  to the Ioffe time $\nu$ gives  the functions 
 \begin{align}
 {\cal I} (\nu, \mu^2) 
&   = 
 \int_{-1}^1 dx 
 \, e^{i x \nu  } \,  f (x, \mu^2)  \    
  \label{ITD}
\end{align}
introduced in Ref.  \cite{Braun:1994jq}  and  called there  the {\it Ioffe-time distributions}. 
In this context, the 
functions  ${\cal M}(\nu , -z^2) $   that are  the  Fourier transforms of
pseudo-PDFs,    should be called the  Ioffe-time {\it pseudo-distributions}
or {\it pseudo-ITDs}.

To get  a relation between the pseudo-PDFs ${\cal P} (x,z_3^2)$  and the
$\overline{\rm MS}$     parton densities $ f(x, \mu^2)$, one can use 
 the  nonlocal  light-cone OPE \cite{Anikin:1978tj,Balitsky:1987bk}
 (see also \cite{Radyushkin:2017lvu}) 
for  the  matrix element defining 
${\cal P} (x,z_3^2)$, i.e., for  the pseudo-ITD. The result  
\begin{align} 
{\cal M } (\nu, -z^2 ) =&  \sum_i  \int_{-1}^1 dw \, C_i (w, z^2 \mu ^2, \alpha_s) \,  {\cal I}_i (w \nu,\mu^2) 
\nn & + {\cal O} (z^2) 
  \  ,
\label{OPE}
\end{align}
has the   structure  similar to that of the usual  OPE  for the DIS structure functions $W (x, Q^2)$.   
In this expression,   the twist-2 coefficient functions $C_i$ are   
given by an expansion in the strong coupling constant $\alpha_s$,
while  ${\cal O} (z^2) $ symbolizes  higher-twist terms.

However, the application of the OPE
to the pseudo-ITDs  and pseudo-PDFs   in QCD  faces 
 complications   related to the gauge link.  Namely, when $z$ is off the light cone,   
the link  generates 
linear $\sim z_3/a$ and logarithmic $\sim \ln (1+z_3^2/a^2)$  ultraviolet (UV) divergences,
where $a$ is an UV regulator with the dimension of length
(it may be a finite lattice  spacing).   
Though  
 disappearing  for $z_3=0$, these   divergences  
require an additional UV  regularization when $z_3$ is finite.

Fortunately, these divergences are multiplicative \mbox{\cite{Polyakov:1980ca,Dotsenko:1979wb,Brandt:1981kf,Aoyama:1981ev,Craigie:1980qs}}  (see also 
 recent Refs. 
 \cite{Ishikawa:2017faj,Ji:2017oey,Green:2017xeu}), and cancel in the ratio,
 the  reduced Ioffe-time distribution,
 \begin{align}
{\mathfrak M} (\nu, z_3^2) \equiv \frac{ {\cal M} (\nu, z_3^2)}{{\cal M} (0, z_3^2)} \   ,
 \label{redm}
\end{align}
 introduced in our paper \cite{Radyushkin:2017cyf}, and partially motivated 
 by this cancellation.
 The remaining $\ln z_3^2$ singularities, present only in the numerator of the ratio,
  are   described by the nonlocal light-cone OPE.

As stated in Ref.  \cite{Orginos:2017kos}, for 
 small spacelike intervals $z^2=-z_3^2$,  and at the leading logarithm level, the reduced pseudo-PDFs are 
related to the $\overline{\rm MS}$ distributions by a simple rescaling of their second arguments, 
namely,  
\begin{align} 
\mu^2  =4e^{-2\gamma_E}/ z_3^2 
  \  ,
\label{Ptof}
\end{align}  
where $\gamma_E$  is the Euler's constant (a more detailed discussion  will be given later on).  
This  rescaling  factor  is very close to 1, since \mbox{$2e^{-\gamma_E}=1.12$. }
However, this factor may be changed by the $ {\cal O} (\alpha_s) $  terms present in the coefficient function.

\section{Structure of one-loop corrections} 
 
 \subsection{Confinement and infrared cut-offs}
 
 There are several standard techniques to calculate gluon radiative corrections in QCD.
 Most of them  are oriented to work in the region of 
 absolute perturbative QCD (pQCD) applicability. 
A
  straightforward use of such methods, however,  
 may need some  care  in applications involving energy scales
 that are not very large.
   For this reason,  let us discuss 
   some features of calculations 
   on the border of applicability of perturbative methods.

To begin with, one should  remember that quarks and gluons are confined, i.e. 
 the  propagators of all   diagrams  (even in a continuum case)
 are embedded in a finite volume whose size is determined by the hadron's radius.  
 The confinement  effects lead, in particular,
  to a rapid   decrease of  correlators like ITDs or pseudo-PDFs  at  distances $z_3$
 larger  than the  hadronic radius  $R$.
Still,  at short distances  one can  use  asymptotic freedom
 and obtain, in particular, the $\ln z_3^2$ singularities.
 
Thus, it makes sense to treat  pseudo-ITDs and  pseudo-PDFs  as  sums  of the soft and hard parts.  
The soft  part  basically 
reflects  the  size of the system  and is assumed to be  finite for $z_3=0$.   The hard part
is   singular for \mbox{$z_3 \to 0$},   and  is   produced by perturbative interactions.
The hard part may be visualized  then as
 generated from the soft part  through a hard exchange kernel $ H (0,z; z_1, z_2)$, 
\begin{align} 
    & {\cal M}^{\rm   hard} (\nu, -z^2=z_3^2)
    \nn & =   \int d^4z_1 \, d^4 z_2 \, H (0,z; z_1, z_2) \,  {\cal M}^{\rm  soft} (z_1,z_2)\  . 
\label{Mhar}
\end{align} 

In the standard  pQCD factorization approaches,
the soft part is mimicked  by on-shell parton states, and the 
\mbox{$\ln z_3^2$-singularities}  appear either as $\ln (z_3^2 m^2)$,
where $m$ is the parton mass or  $\ln (z_3^2 \mu_{\rm IR}^2)$, 
where $\mu_{\rm IR}$ is the scale used in dimensional regularization 
of infrared singularities in  the  case of massless partons.

Since   ${\cal  M}^{\rm  soft} (z_1,z_2)$ in Eq. (\ref{Mhar}) 
rapidly decreases for large separations
$|z_1-z_2|$,  the hadronic size $R$  provides an infrared cut-off for the integral,
even when the quarks are massless.  While at   short distances   one   gets  the
 $\ln (z_3^2 / R^2)$ behavior,   the logarithmic form  is just an approximation
valid for $z_3 \ll R$. Such 
a restriction may be hard to implement on the lattice.

 Of course, the exact form of the IR regularization  imposed  
by confinement is not known. To get  a feeling,
  let us take an infrared regularization by a  mass term.
 A typical integral producing the  $\ln z_3^2$  singularity then  has the form
   \begin{align} 
  I_K (z_3^2) = \int_0^\infty \frac{d\alpha}{\alpha} e^{-z_3^2/4\alpha -   \alpha m^2} \ ,
  \label{IK}
     \end{align} 
where $\alpha$ is the Schwinger's $\alpha$-parameter and $m$ is the infrared  regulator.  
 One can see that
   \begin{align} 
  I_K (z_3^2) = &2 K_0 (mz_3) \nn & = - \ln (m^2 z_3^2) + 2 \ln (2e^{-  \gamma_E })+ {\cal O} (z_3^2)  \  ,
  \label{Kexp} 
     \end{align} 
where $K_0(mz_3)$ is the modified Bessel function.  Its expansion for small $z_3$
explicitly  shows the expected $\ln (z_3^2 m^2)$ singularity.

The usual pQCD factorization   procedure   is to split 
$\ln (z_3/ R)$ into the short-distance part $\ln (z_3 \mu)$   that is attributed  to  
the coefficient function  and the long-distance part $\ln ( 1/ \mu R )$
that is   absorbed into the ``renormalized'' PDF $f(x,\mu^2)$.
Given the commonly used   lattice   spacing $a\sim 0.1$ fm  and the hadron size
$R \lesssim 1$ fm, the question is whether  there is enough interval 
for the logarithmic part 
of the \mbox{$z_3$-dependence} to   be visible  in the  data  at all.

An important feature of the Bessel  function $K_0 (mz_3)$ is that it 
  exponentially decreases when 
$z_3$  exceeds the infrared cut-off $1/m$. 
Thus, if   instead of the  short-distance approximation of  
$I_K(z_3^2)$ by $\ln (1/z_3^2)$, 
one would use the ``exact'' $I_K(z_3^2)$ function  
for the evolution term, 
 there will be no evolution corrections for large $z_3$.
In other words, the logarithmic 
evolution disappears at large distances.

\subsection{Rescaling relation} 

To fix a  relation between the pseudo-PDF scale $z_3$  and the $\overline{\rm MS}$ scale $\mu$,
one should take into account constant terms, like $2 \ln (2e^{-  \gamma_E })$  in Eq. (\ref{Kexp}).  
In the $\overline{\rm MS}$-OPE approach,
  one takes $z^2=0$ 
and  then  applies 
 the dimensional regularization which   adds  the  $\alpha^{\epsilon}$ factor into the integral
(\ref{IK}) making it convergent. After that, one    uses the $\overline{\rm MS}$-prescription, 
which is   arranged  to produce  exactly
 $\ln (\mu^2/m^2)$ as the result in this case, 
    \begin{align} 
  I_{Dm} (\mu^2) =& \int_0^\infty \frac{d\alpha}{\alpha}  (\alpha \mu^2 e^{\gamma_E})^{\epsilon}  e^{-   \alpha m^2} \,
  =\Gamma (\epsilon) \left ( \frac{\mu^2 e^{\gamma_E}}{m^2} \right )^\epsilon \nn & \to 
  \frac1{\epsilon}   + \ln (\mu^2/m^2)  \  . 
  \label{ImM}
     \end{align} 
 Thus,  the constant  term in Eq. (\ref{Kexp}) 
provides the leading-logarithm rescaling coefficient 
$2e^{-  \gamma_E}$  between the pseudo-PDFs and $\overline{\rm MS}$ parton distributions
expressed by Eq. (\ref{Ptof}).

One may ask what happens  if one uses another type of    the IR regularization.
In particular,  the  Gaussian models for TMDs suggest that the 
decrease for large  $z_3$ is also Gaussian.  One may expect  
 that    the hard correction
should resemble,  for  large $z_3$,  the behavior of the soft part.  
  Thus,   the exponential $e^{-m|z_3|}$ fall-off of the modified 
Bessel function may  look   too slow. 
A  Gaussian decrease can  be easily  provided by a sharp IR cut-off  
    \begin{align} 
  I_G (z_3^2) = \int_0^{z_0^2/4}  \frac{d\alpha}{\alpha}    e^{-z_3^2/4\alpha } = \Gamma [0,z_3^2/z_0^2]
  \label{IDexp}
     \end{align} 
applied to  Eq. (\ref{IK}).  For small $z_3^2$,  the incomplete 
gamma-function  $ \Gamma (0,z_3^2/z_3^2)$   has a logarithmic singularity
   \begin{align} 
  \Gamma (0,z_3^2/z_0^2) =  \ln (z_0^2/z_3^2) -  \gamma_E + {\cal O} (z_3^2)  \  ,
  \label{Dexp0}
     \end{align} 
     while for large $z_3^2$, the function  $I_G (z_3^2)$  has  a Gaussian  $e^{-z_3^2/z_0^2}$  fall-off.  
Again, we  can  calculate the $z_3=0$ version of 
Eq. (\ref{IDexp})  using the $\overline{\rm MS}$-scheme to obtain 
  \begin{align} 
  I_{DG} (\mu^2) = & \int_0^{z_0^2/4}  \frac{d\alpha}{\alpha}  (\alpha \mu^2  e^{\gamma_E})^{\epsilon}    = 
  \frac1{\epsilon} \,  \left (\frac{ z_0^2  \mu^2  e^{\gamma_E}}{4}  \right )^{\epsilon} \nn & 
  \to  \frac1{\epsilon} \, + \ln (z_0^2 \mu^2    ) -2  \ln ( 2 e^{-\gamma_E}) -  \gamma_E \ . 
  \label{IDexp2}
     \end{align} 
One can see that the pseudo-PDF/PDF rescaling (\ref{Ptof}) remains intact. 
This   is a natural result, because the   relation between the finite-$z_3$ 
and $\overline{\rm MS}$  cut-offs  concerns only the short-distance  properties 
of  the bilocal operator.

\subsection{One-loop correction} 
 
The  discussion given in the  previous section   addresses  only the 
overall  rescaling between  two regularization schemes
(just like    the  relation between the values of the QCD scale $\Lambda$ in, say, MOM and
$\overline{\rm MS}$ schemes). 
To  establish a connection  between the pseudo-PDFs and  the 
$\overline{\rm MS}$-PDFs, we  need, in addition, the constant part of the one-loop 
coefficient function in the nonlocal OPE of  \mbox{Eq. (\ref{OPE}).  } 
It   was given  in Refs. \cite{Ji:2017rah} and \cite{Radyushkin:2017lvu},
with some differences between them. 
After rechecking our calculation and fixing  typos,
we present our result  in the form 
  \begin{align} 
  &
{\mathfrak M}  (\nu, z_3^2)    = {\mathfrak  M}^{\rm soft}  ( \nu, 0)  -  \frac{\alpha_s}{2\pi} \, C_F
\int_0^1  dw \,  \left \{   \frac{1+w^2} {1-w}    \right.  \nn & \left.  \hspace{2cm} \times
  \left [  \ln \left (z_3^2m^2\frac{  e^{2\gamma_E}}{4} \right )   +1 \right ]  
\right. \nn    & \left. 
+ 4  \,  \frac{\ln (1-w)}{1-w} 
\right \}  \left [
{\mathfrak  M}^{\rm soft}  (w \nu,0) -  {\mathfrak  M}^{\rm soft}  ( \nu,0) \right ]\  .
 \label{Mhard} 
 \end{align}

Turning to the PDF counterpart, we take $z^2=0$ and using the  $\overline{\rm MS}$ scheme for the UV divergence, obtain 
   \begin{align}
{\cal I}  (\nu, \mu^2)
  =&  {\mathfrak  M}^{\rm soft}  ( \nu, 0) \nn & -  \frac{\alpha_s}{2\pi} \, C_F\, 
    \int_0^1  dw \,   
  \left [
{\mathfrak  M}^{\rm soft}  (w \nu, 0) -  {\mathfrak  M}^{\rm soft}  ( \nu, 0) \right ]\nn & 
  \nn & \times \left \{    \frac{1+w^2} {1-w} \,  \ln (m^2/ \mu^2)   + 2 (1-w) \right \}
 .
\label{Msbar}
 \end{align}
 
The logarithmic part here involves a convolution that  may be symbolically written as  
   $
B \otimes {\mathfrak M}\,  (\nu) $ where 
     \begin{align}
B(w) = 
\left [ \frac{1+w^2}{1- w} \, \right ]_+  
\label{AP}
 \end{align}
 is  the Altarelli-Parisi kernel  \cite{Altarelli:1977zs}. 
 
Combining Eqs. (\ref {Mhard}) and (\ref{Msbar}), we obtain 
 the relation
  \begin{align} 
{\cal I}  (\nu, \mu^2)    = & {\mathfrak  M} ( \nu, z_3^2) +  \frac{\alpha_s}{2\pi} \, C_F\,  
\int_0^1  dw \,   {\mathfrak  M}  (w \nu,z_3^2)   \nn &
 \times  \left  \{   B(w) \, \left [ \ln \left (z_3^2\mu^2 \frac{  e^{2\gamma_E}}{4} \right )  +1 
 \right ]   \right. \nn & \left. 
+   \left [ 4  \frac{\ln (1-w)}{1-w} - 2 (1-w)  \right ]_+ \right  \} 
\  
\label{MNL}
 \end{align}
 which is in agreement with a recent result of Ref. \cite{Izubuchi:2018srq}
 (see  also  Ref. \cite{Zhang:2018ggy}).
Eq. (\ref{MNL})   allows  one to convert the  data points 
 for ${\cal  M} ( \nu, z_3^2) $  into the ``data'' for ${\cal  I} (\nu, \mu^2)$.

 The first contribution in   the   second line is  an obvious  term reflecting 
 the  general multiplicative scale  difference between  the $z^2$ and $\overline{\rm MS}$  cut-offs. 
If  all  the further  terms  are  neglected, then the only difference between
 $ {\mathfrak  M} ( \nu, z_3^2) $  and ${\cal  I} (\nu, \mu^2)  $  
 is  just the rescaling \mbox{$\mu^2  =4e^{-2\gamma_E}/ z_3^2$.}   
 In that case,   one can evolve the $ {\mathfrak  M} ( \nu, z_3^2) $  data to a particular 
 $z_3$   value  $z_0$, and treat (in this approximation) the resulting function 
 $ {\mathfrak  M} ( \nu, z_0^2) $ as the  $\overline{\rm MS}$  ITD 
 corresponding to the scale \mbox{$\mu = 2 e^{-\gamma_E}/ z_0$,}
 which is numerically close to $1/z_0$. 
 
 This   simple rescaling relation (used in Ref. \cite{Orginos:2017kos})    is modified 
 when  the   further  terms  of Eq. (\ref{MNL})  are included.
 In particular, 
 the  term  proportional
to the Altarelli-Parisi kernel $B(w)$ may be absorbed into the $\ln z_3^2$ term,
which would just  change the  rescaling relation 
  into   $\mu = 2 e^{-1/2-\gamma_E}/ z_0$.

    \begin{figure}[t]
    \centerline{\includegraphics[width=3in]{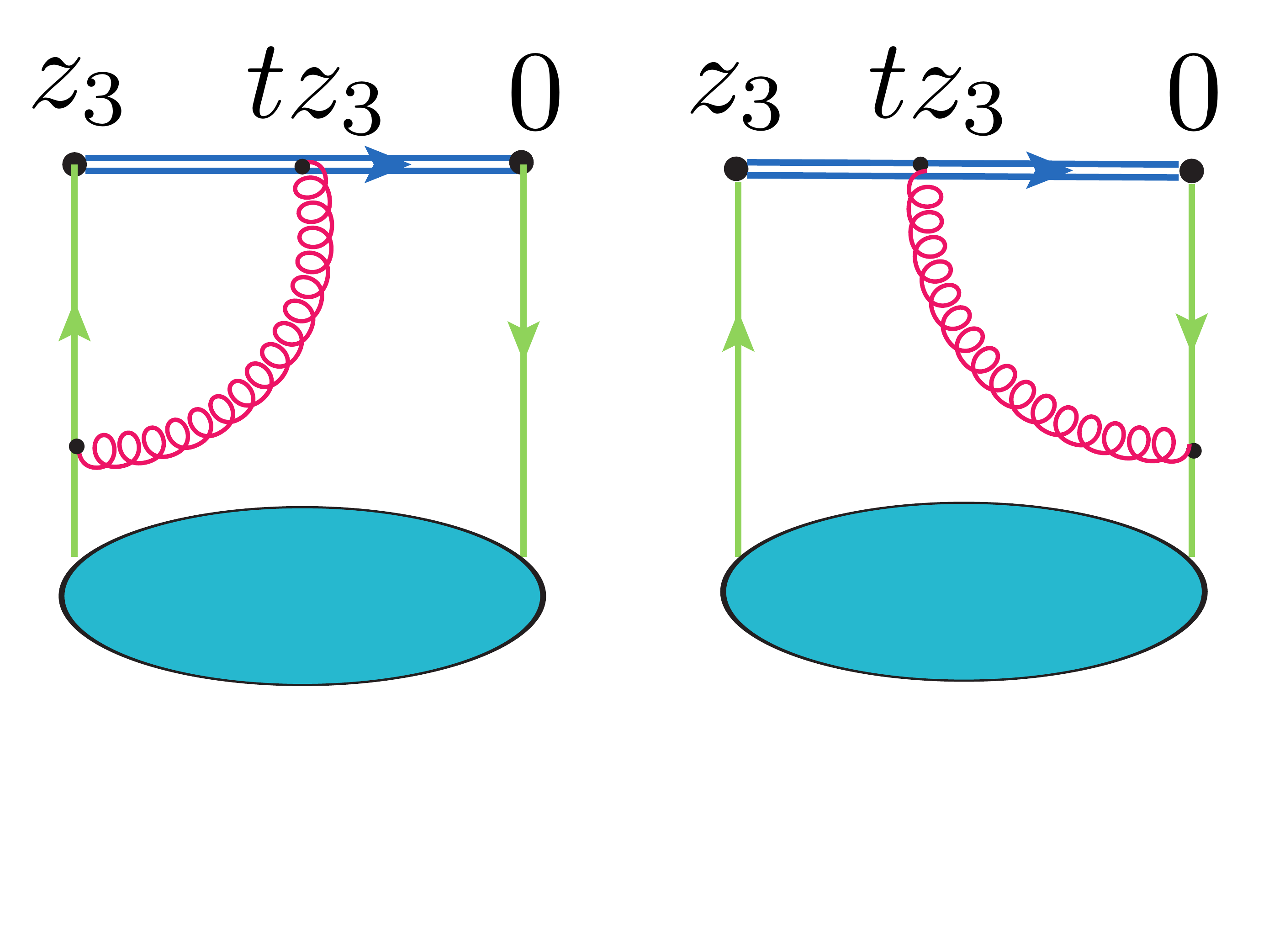}}
    \vspace{-1.2cm}
    \caption{Coordinate representation for  diagrams  producing a  large one-loop correction. 
        \label{link}}
    \end{figure}

The term with $[\ln (1-w)]/(1-w)$  produces  a large negative contribution.    
In Feynman gauge, according to Ref. \cite{Radyushkin:2017lvu},  it comes from the evolution part of the  vertex diagrams 
  involving
    the gauge link      (see Fig.\ref{link}).  The key point is   that the gluon is attached there  to 
    a running $tz_3$  position on the link.  After integration
    over $t$,  etc., the net outcome is  that the $z_3$-dependence 
    of these diagrams  is generated by an effective  scale  smaller than $z_3$.    
    Indeed, let  us   combine the  $[\ln (1-w)]/(1-w)$   term with the $\ln z_3^2$ logarithm
    by rewriting Eq. (\ref{MNL}) as 
    \begin{align} 
{\cal I}  (\nu, \mu^2)    = & {\mathfrak  M} ( \nu, z_3^2) +  \frac{\alpha_s}{\pi} \, C_F\,  
\int_0^1  dw \,   {\mathfrak  M}  (w \nu,z_3^2)   \nn &
 \times  \Biggl   \{   \frac{1+w^2}{1-w}  \,  \ln \left [(1-w)z_3  \mu \frac{  e^{\gamma_E+1/2}}{2} \right ]
\nn & 
+   \left [   (w+1)  \ln (1-w)  -(1-w) \right ] \Biggr  \} _+ 
\   . 
\label{MNL2}
 \end{align}
 We  see that $z_3$  enters now through  a running $(1-w)z_3$  location. 
 The remaining  $(w+1) \ln (1-w)$  term is much less  singular 
 than $B(w)$ for $w=1$, and does not produce large contributions. 
 
 Thus, the magnitude 
 of the one-loop correction is governed by the combined evolution logarithm.
 It cannot be made zero by a particular choice of $\mu$  because
 it depends on the integration variable $w$.
 Still,   the $w$-integrated contribution will vanish for some $\mu$   that 
 we may write as 
     \begin{align} 
\mu =  \frac{2 e^{-1/2-\gamma_E}}{\langle 1-w \rangle   }\, \frac1{z_3} \ , 
\end{align}
    where $\langle 1-w \rangle $  is the ``average''  value of $1-w$.
    Since $B(w)$  is strongly enhanced for $w=1$, we should expect
    that $\langle 1-w \rangle $ is  numerically small,
    leading to a $\mu \sim k/z_3$ rescaling with a  rather large coefficient $k$.
 As we will see, $k \sim 4$ in this case.

  Again, one may  ask  if 
the perturbative formula (\ref{MNL})  involving the 
$\ln  z_3^2 $ logarithm may be applied to  actual   lattice data. 
 In particular, our exercise with the mass-term IR regularization 
 and the  resulting  Bessel function shows 
 that 
 the logarithmic behavior $\ln z_3^2 $ 
of the hard term is valid   only for $z_3 $  values well   below 
the IR cut-off $R$, which is given by  the hadron size in our case. 
Hence, a practical question is whether 
  the data  really 
show a logarithmic evolution behavior  in some region of  \mbox{small $z_3$. }

\section{Evolution in   lattice data}

\subsection{General features} 
  
 An exploratory  lattice study 
 of the reduced pseudo-ITD  ${\mathfrak M} (\nu, z_3^2)$
 for the valence $u_v-d_v$ parton distribution in the nucleon has  been reported in Ref.
 \cite{Orginos:2017kos}.  
 An amazing observation made there was that, 
 when plotted as functions of $\nu$, the data both for real
 and imaginary  parts lie close to respective universal curves.
 The data  
 show no polynomial $z_3$-dependence for large $z_3$.
 Given that $z_3^2/a^2$ changes in the explored  range
 from 1 to about 200, we  interpret this result as the  {\it  total 
 absence}  of higher-twist terms in the reduced
 pseudo-ITD.
 
 As  explained in Refs.  \cite{Radyushkin:2017cyf,Orginos:2017kos}   and in the Introduction, 
 such an outcome corresponds to  a factorization  
  of the $\nu$-  and $z_3^2$-dependences of the soft part of the    
 Ioffe-time distribution ${\cal  M} (\nu, z_3^2)= M (\nu) {\cal  M} (0, z_3^2)$.  
 In terms of TMD  ${\cal F} (x, k_\perp^2)$, this corresponds to factorization   
 of its  $x$- and $k_\perp^2$-dependences   in the region of soft   $k_\perp$. 
 However, as observed in Ref.  \cite{Orginos:2017kos}, there is quite visible $z_3$-dependence 
 for small values of $z_3$, namely, $z_3 \lesssim 6a$, that may be explained by perturbative evolution.  

 Let us  consider first the real part.  
 It  corresponds  to 
  the cosine Fourier transform 
  \begin{align} 
   {\mathfrak  R} (\nu)  \equiv 
{\rm Re} \,  {\mathfrak  M} (\nu)=  & \int_0^1 dx \,  
\cos (\nu x)  \, q_v  (x)  
\label{MC}
     \end{align}
of the function  $q_v(x)$  corresponding to  the valence combination, i.e., the 
difference  $q_v(x) = q(x) -\bar q(x)$  of 
quark and antiquark   distributions. 
 In our case,   $q= u-d$.

    \begin{figure}[b]
    \centerline{\includegraphics[width=3.3in]{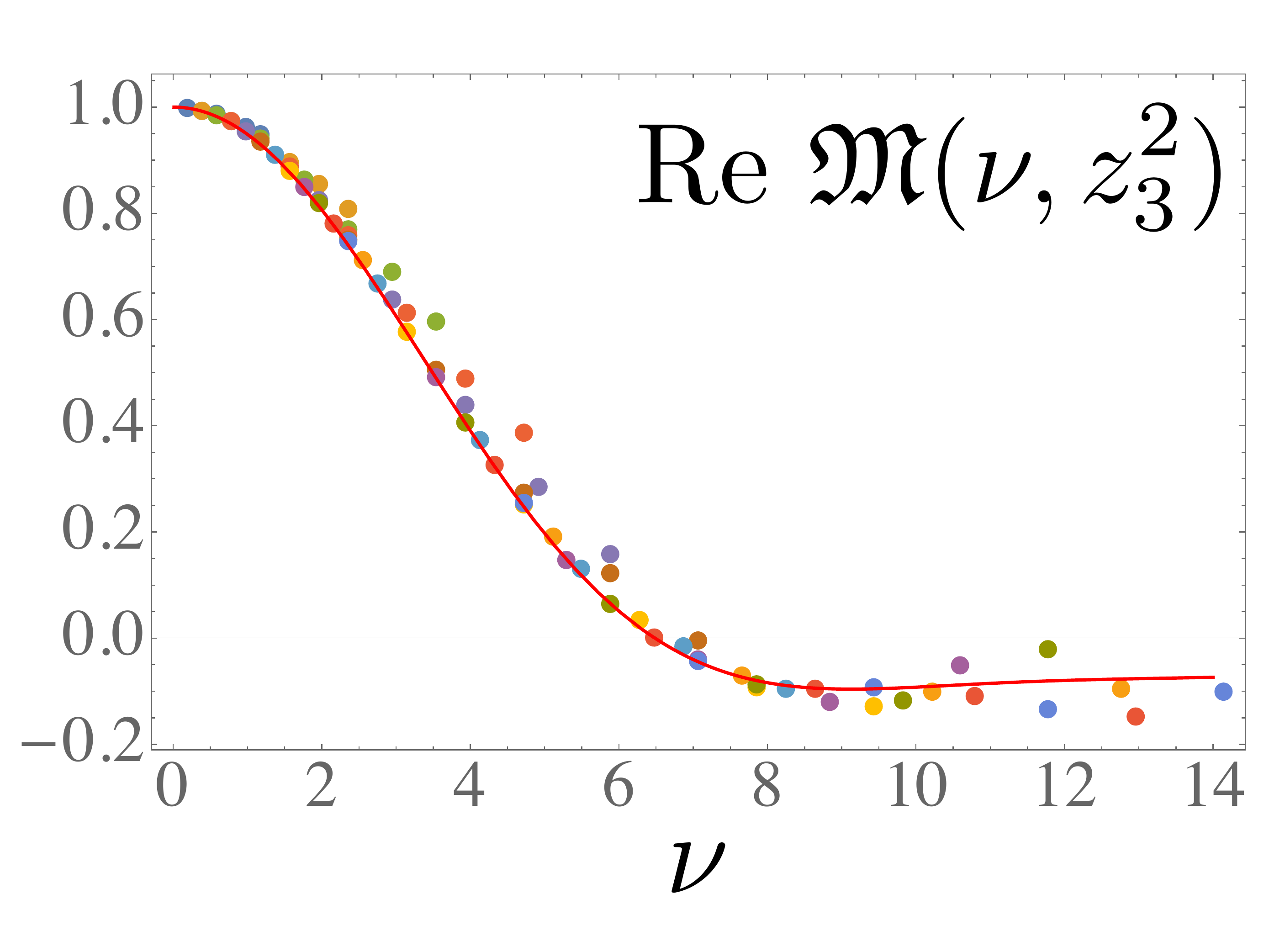}}
    \caption{Real  part of ${\mathfrak  M} (\nu, z_3^2)$  plotted as a function
    of $\nu =Pz_3$ and 
    compared to the curve given by Eqs. (\relax {\ref{MC}}), (\relax {\ref{qV}}).
        \label{realc}}
    \end{figure}

In  Ref. \cite{Orginos:2017kos},  it was  found that the   
data for the real part are  very close  (see Fig. \ref{realc})  to the curve
${\mathfrak R}_f ( \nu)$ 
generated  by    the function 
     \begin{align} 
    f(x) =  \frac{315}{32} \sqrt{x}  (1-x)^{3}\,  . 
    \label{qV}
    \end{align}
This shape  was obtained by   forming  cosine Fourier transforms  of the 
normalized $x^a(1-x)^b$-type  functions 
and fixing  the parameters $a,b$  through   fitting    the  data. 

   \begin{figure}[t]
  \centerline{\includegraphics[width=3.3in]{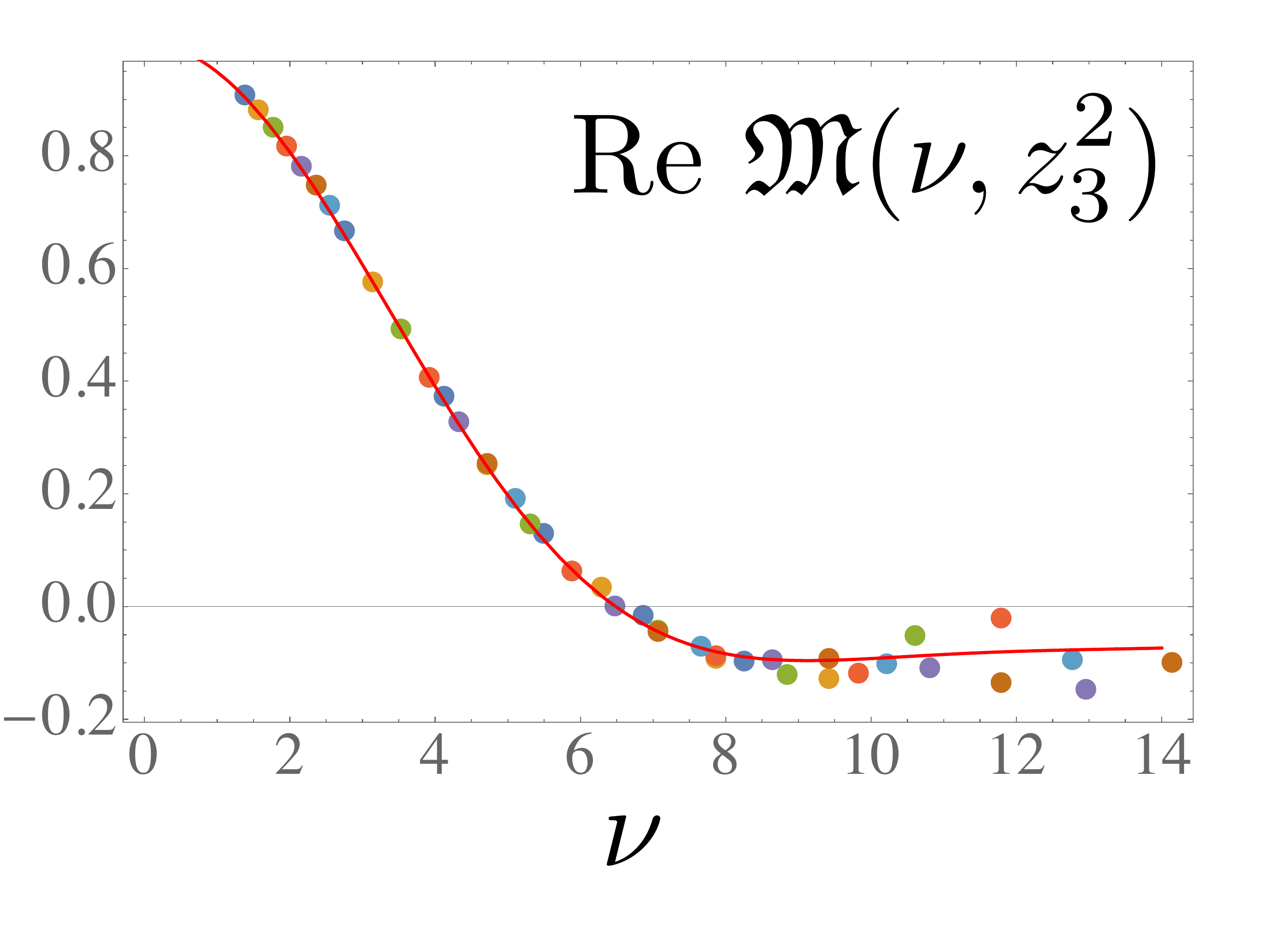}}
    \caption{Real part   of ${\mathfrak  M} (\nu, z_3^2)$  for  $z_3$  ranging  from  $7a$ to  $13a$.
    \label{M713}}
    \end{figure}

\begin{figure}[t]
  \centerline{\includegraphics[width=3.3in]{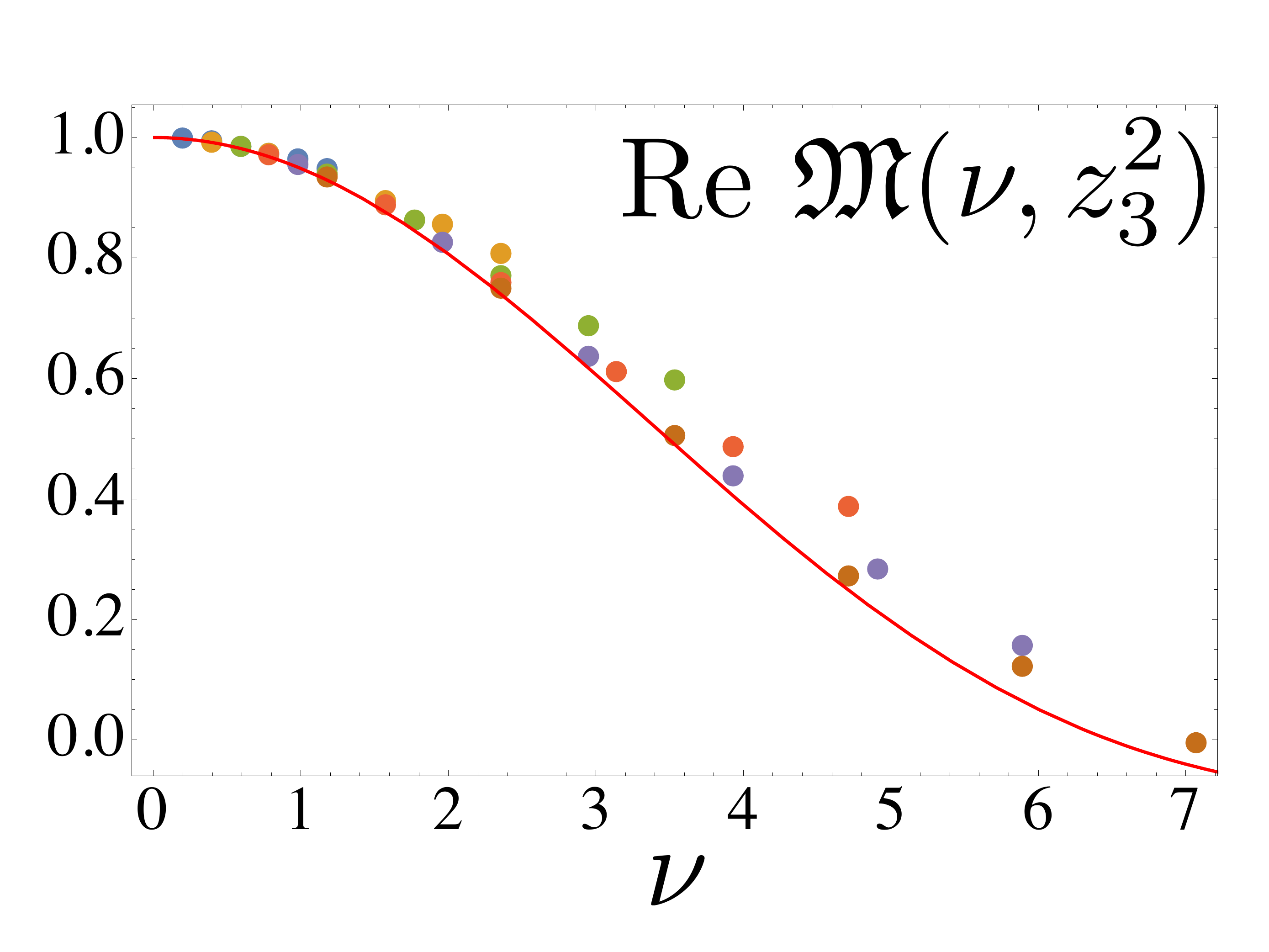}}
    \caption{Real part  of ${\mathfrak  M} (\nu, z_3^2)$ for $z_3$ ranging   from  $a$ to  $6a$. 
     \label{M16}}
    \end{figure}

While 
 all the data points have been  used in the fit, 
  the shape of the curve   is obviously dominated by  
the points with  smaller values of  \mbox{ Re\ ${\mathfrak  M} (\nu, z_3^2)$.  }
To give a more detailed illustration, we show in Fig. \ref{M713} the points 
corresponding to $z_3$  values in the range $7a \leq z_3 \leq 13a$. 
As  one can see,  there  is some scatter for the points 
with the largest values of $\nu$ in the region $\nu \gtrsim 10$,
where  the finite-volume effects  become important. 
Otherwise,  practically all the points lie on the universal 
curve based on $f(x)$. 
 In this sense, there is no $z_3$-evolution visible  in the large-$z_3$ data.

In Fig. \ref{M16}, we show the points in the region  \mbox{$a \leq z_3 \leq 6a$} 
(note that,   on the lattice,  $z_3=0$ means that also $\nu=0$,
and ${\mathfrak  M} (0, 0)=1$ by definition). 
 In this case, all   the points lie  
higher than  the universal curve.  
We recall that the perturbative evolution  
increases   the real part of the pseudo-ITD
when $z_3$ decreases. Thus,  one may conjecture that
the  observed higher values 
of ${\mathfrak R}$ for smaller-$z_3$ points may be a consequence of the evolution.

\begin{figure}[t]
  \centerline{\includegraphics[width=3.3in]{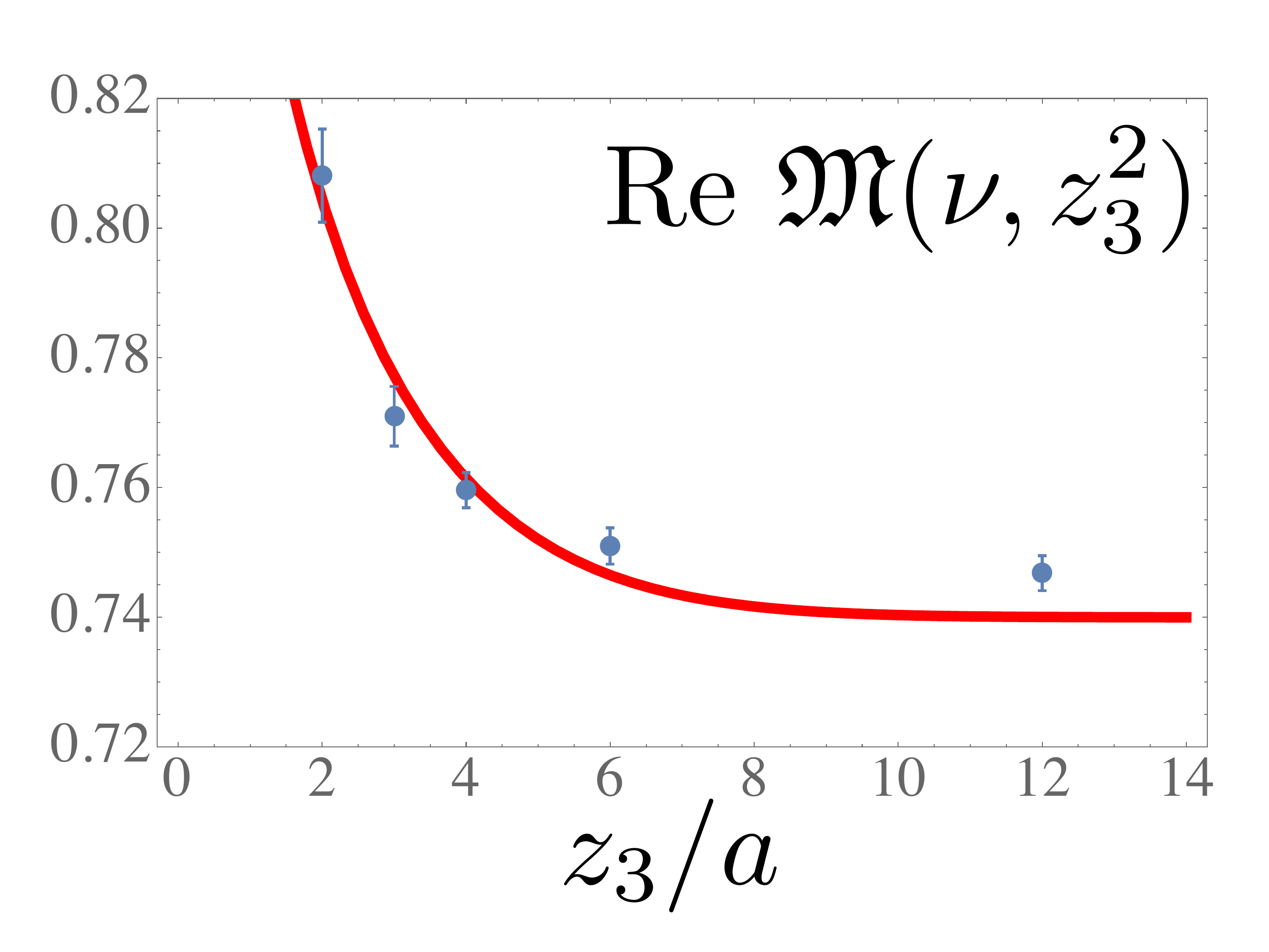}}
    \caption{Dependence  on $z_3$ for $\nu=3\pi/4\approx2.3562$. 
    \label{nu23}}
    \end{figure}

A  typical pattern of the $z_3$-dependence of the lattice points 
is shown in \mbox{Fig. \ref{nu23}} 
for a ``magic''  Ioffe-time value $\nu =3\pi/4$ that may be obtained from 
five different combinations of $z_3$ and $P$ values 
used in Ref. \cite{Orginos:2017kos}.  The shape of the eye-ball fit  line is
given by the incomplete gamma-function $\Gamma (0,z_3^2/30a^2)$.
This function entirely conforms
to the expectation that the  $z_3$-dependence 
has   a ``perturbative'' logarithmic $\ln (1/z_3^2)$  behaviour  for small $z_3$,
and 
rapidly vanishes  for $z_3$ larger than $6a$.

As expected,  
${\mathfrak R} ( \nu, z_3^2)$  decreases    when $z_3$ increases.
We also see that  the evolution ``stops'' for large $z_3$.
In this context, the overall  curve  based on Eq.  (\ref{qV}) corresponds to  the ``low normalization point'', 
i.e.,  to   the region, where the perturbative evolution is absent. 

\subsection{Building $\overline{\rm MS}$  ITD}

 Thus, we see   that the data of \mbox{Fig. \ref{nu23}}  
show a logarithmic evolution behavior  in  the small  $z_3$ region.  
Still,    the 
\mbox{$z_3$-behavior }  starts to visibly deviate 
from a pure logarithmic $\ln z_3^2$  pattern for $z_3\gtrsim 5a$.     
This  sets the  boundary $z_3 \leq 4a$  on the ``logarithmic region''.
So, let us try to  use Eq. (\ref{MNL}) in that region to construct 
the $\overline{\rm MS}$ ITD.

It is instructive to split the contributions in Eq.  (\ref{MNL}), where we will 
denote  ${\rm Re} \, {\cal I}  (\nu, \mu^2)  \equiv  {\cal I}_R  (\nu, \mu^2) $.
The first, ``evolution'' part, given by
  \begin{align} 
 {\cal I}^{\rm ev} _R  (\nu, \mu^2)    =&  {\mathfrak  R} ( \nu, z_3^2) +  \frac{\alpha_s}{2\pi} \, C_F\,  
\int_0^1  dw \,   {\mathfrak  R}  (w \nu,z_3^2)   \nn &
 \times    B(w) \,   \ln \left (z_3^2\mu^2 \frac{  e^{2\gamma_E}}{4} \right )  
\  
\label{MNLev}
 \end{align}
(recall that  $ {\mathfrak  R} ( \nu, z_3^2) \equiv {\rm Re} \,  {\mathfrak  M} ( \nu, z_3^2$)) 
corresponds to the leading logarithm approximation used in Ref. \cite{Orginos:2017kos}.
 For $z_3 =2 e^{-\gamma_E}/\mu$, the logarithm vanishes, and  we have
   \begin{align} 
 {\cal I}^{\rm ev}_R   (\nu, \mu^2)    =&  {\mathfrak  R} ( \nu, (2 e^{-\gamma_E}/\mu)^2) =
  {\mathfrak  R} ( \nu, (1.12/\mu)^2) \  . 
\label{IMev}
 \end{align}

 This  happens, of course, only  if,   for an appropriately chosen $\alpha_s$,  
 the \mbox{$ \ln  z_3^2$-dependence}  
 of the one-loop correction  cancels  the actual $z_3^2$-dependence of the data,
 visible as scatter in  the data points in Fig. \ref{M16}.  
  In Ref. \cite{Orginos:2017kos}, it was found 
 that this happens when  $\alpha_s/\pi \approx 0.1$. 
 Thus,  Eq. (\ref{MNL})  is accurate  only in the  region,
 where the data show a {\it  logarithmic} dependence on $z_3$, i.e., $z_3 \leq 4a$ 
 in our case.

\begin{figure}[t]
  \centerline{\includegraphics[width=3.4in]{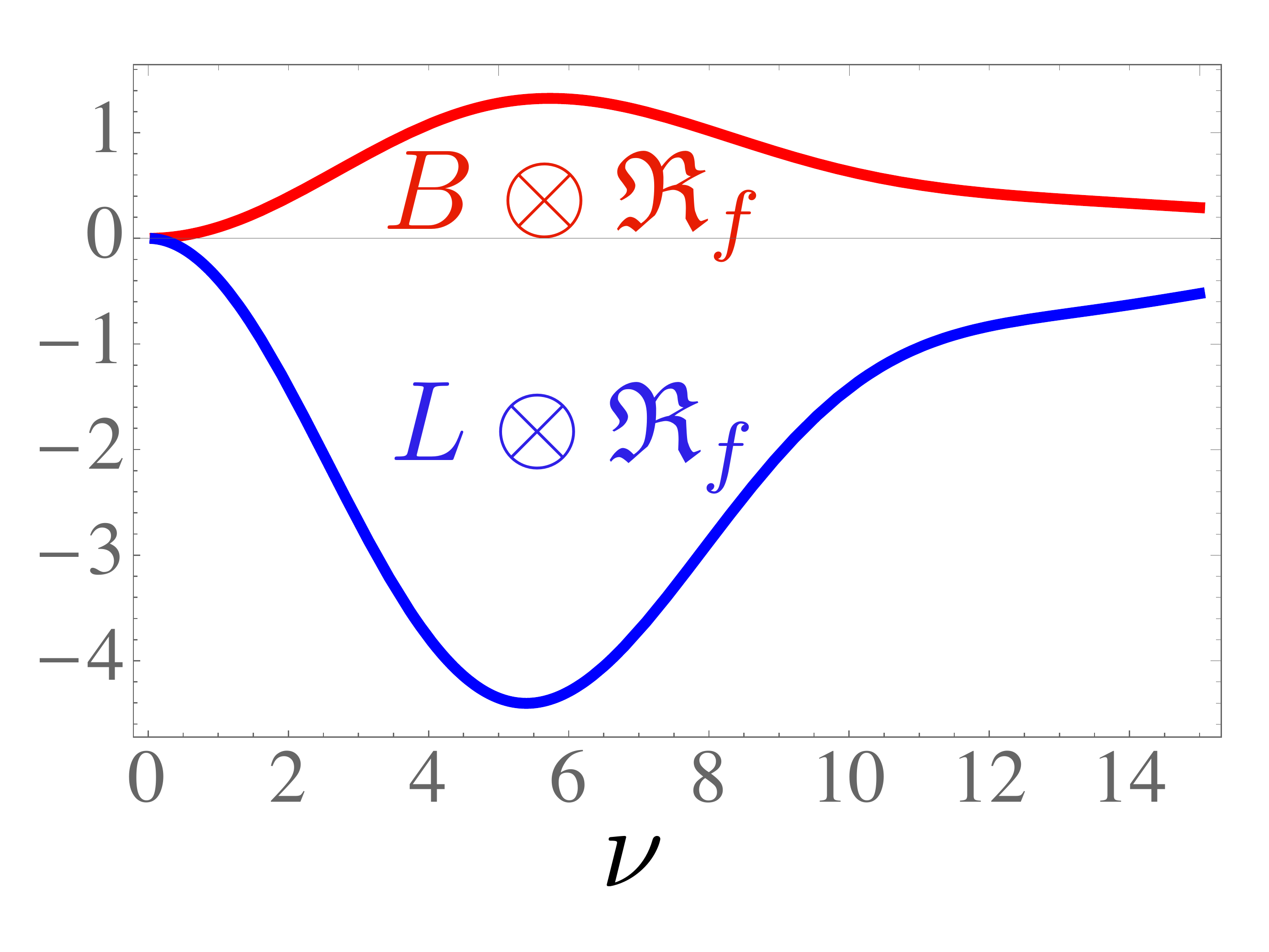}}
    \caption{Functions $B \otimes  {\mathfrak R}_f$ (upper line) and $L \otimes  {\mathfrak R}_f$
   (lower line)  of Eq. (\ref{MNLo}).
    \label{BMLM}}
    \end{figure}

 Since the difference between 
  ${\mathfrak  R}  (w \nu,z_3^2)$  and ${\mathfrak R}_f (w \nu)$
  is ${\cal O} (\alpha_s)$, we may   replace 
  ${\mathfrak  R}  (w \nu,z_3^2)$ by    ${\mathfrak R}_f (w\nu)$ in Eq. (\ref{MNLev})
  (recall that ${\mathfrak R}_f (\nu)$ 
 corresponds  to the  PDF of  Eq. (\ref{qV})).
 The remaining part of   ${\cal I}  (\nu, \mu^2) $ 
 (where we have already substituted  ${\mathfrak  R}  (w \nu,z_3^2)$  by  ${\mathfrak R}_f (w \nu)$)
    \begin{align} 
{\cal I}^{\rm NL}_R   (\nu)    = & \frac{\alpha_s}{2\pi} \, C_F\,  
\int_0^1  dw \,   {\mathfrak R}_f (w \nu)   \nn &
 \times  \left  \{   B(w) \, 
+   \left [4\,   \frac{\ln (1-w)}{1-w} -2 (1-w)  \right ]_+ \right  \} 
 \nn & \equiv  \frac{\alpha_s}{2\pi} \, C_F\, \left  [ B \otimes  {\mathfrak R}_f + L \otimes  {\mathfrak R}_f\right ]
\  
\label{MNLo}
 \end{align}
is due to  corrections 
 beyond the leading logarithm approximation.

 As  we have discussed, the $ L \otimes  {\mathfrak R}_f$  term 
 reflects the fact that  the actual scale in the evolution part 
 of the vertex diagrams is less than $z_3$. 
 To illustrate its impact, 
  we show,      in Fig. \ref{BMLM},  the  functions  $B \otimes  {\mathfrak R}_f$ and $L \otimes  {\mathfrak R}_f$. 
  One can see that the last one is negative and rather  large.   
  Its $\nu$-dependence is similar to that of the $B \otimes  {\mathfrak R}_f$ function. 
  In fact, 
 in the $\nu <5$ region, we have  $L \otimes  {\mathfrak R}_f \approx -3.5 B \otimes  {\mathfrak  R}_v$.
 Thus, the combined effect of these two terms is close to that of $-2.5 B \otimes  {\mathfrak R}_f$.
 As a result, the inclusion of these terms may be approximately treated as a LLA evolution 
 with a  modified rescaling factor. Specifically,
 we may write
    \begin{align} 
 {\cal I}_R   (\nu, \mu^2)    \approx &  \, {\mathfrak  R} ( \nu, (2 e^{1.25-\gamma_E}/\mu)^2)
   \approx   \, {\mathfrak  R} ( \nu, (4/\mu)^2)   \  . 
\label{IMevL}
 \end{align}
 Thus, the  rescaling factor has  changed by a  factor of 4 compared to the  
 original LLA value!

We may use $\mu \approx 4/z_3$  as a guide, but
 the  actual  numerical   calculations should, of course,  be  done  using the ``exact'' Eq. (\ref{MNL}).
To proceed, 
we choose the value \mbox{$\mu=1/a$}  which,
 at the  lattice spacing of $0.093$ fm used in Ref. \cite{Orginos:2017kos} 
  is approximately  \mbox{2.15 GeV.}  
   The  estimate (\ref{IMevL})  tells us  that 
  the   ITD  $ {\cal I}_R   (\nu, \mu^2)$    at this scale should be close to the pseudo-ITD
  ${\mathfrak  R} ( \nu, z_3^2) $
   for  \mbox{$z_3\approx 4a$,}  a distance  that is on the border of  the $z_3 \leq 4a$   region. 
Taking   the  value
  $\alpha_s/\pi =0.1$ used in Ref. \cite{Orginos:2017kos}  and applying  the full one-loop relation (\ref{MNL}) 
to the data with  $z_3 \leq 4a$, we generate   the  points for   ${\cal I}_R  (\nu, (1/a)^2) $. 

 \begin{figure}[t]
  \centerline{\includegraphics[width=3.35in]{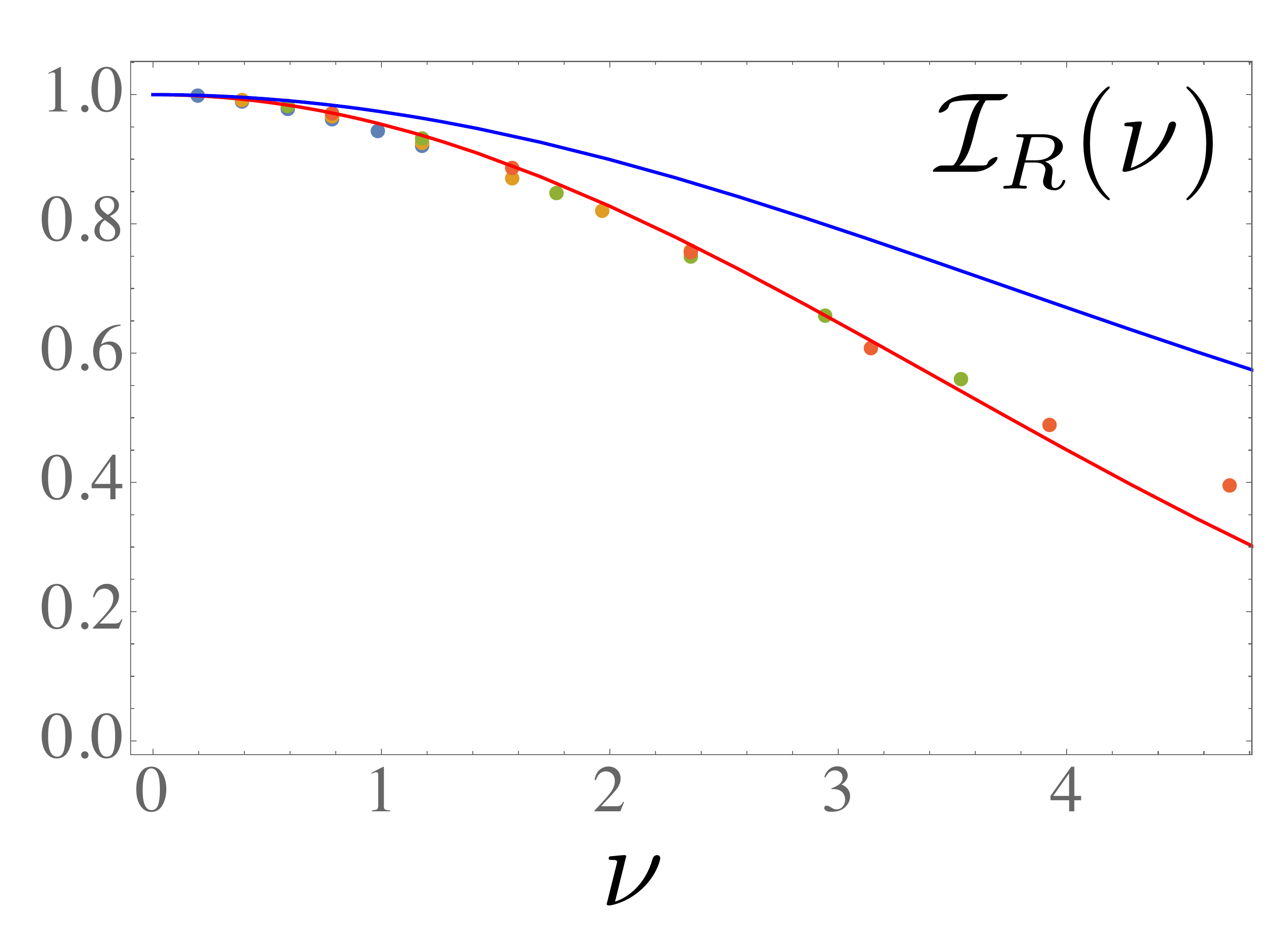}}
    \caption{Function  ${\cal I}_R  (\nu, \mu^2) $  for
    $\mu = 1/a$  calculated using  the data with   $z_3$   from  $a$ to  $4a$.
        The upper curve corresponds to the ITD of the CJ15 global fit PDF.
    \label{Msbar16}}
    \end{figure}

  As seen from   \mbox{Fig. \ref{Msbar16},}  
all the  points are close to  some  universal    curve
   with a rather small scatter.  
     The  curve itself  
     was obtained by fitting the points by 
 the cosine transform  of a normalized $N x^a (1-x)^b$ distribution, 
which gave   
   $a=0.35$ and $b=3$.
   The magnitude of the scatter illustrates  the error of the fit for the 
   ITD in the $\nu \leq 4$ region. 
    In Fig. \ref{Msbar16},  we compare 
  our $\mu=1/a$ ITD  with the ITD  
 obtained from the global fit PDFs   
 corresponding to the CJ15   \cite{Accardi:2016qay}  global fit.  
 One can see that our ITD is systematically below the curve based on the global fit PDFs.  

\begin{figure}[t]
  \centerline{\includegraphics[width=3.6in]{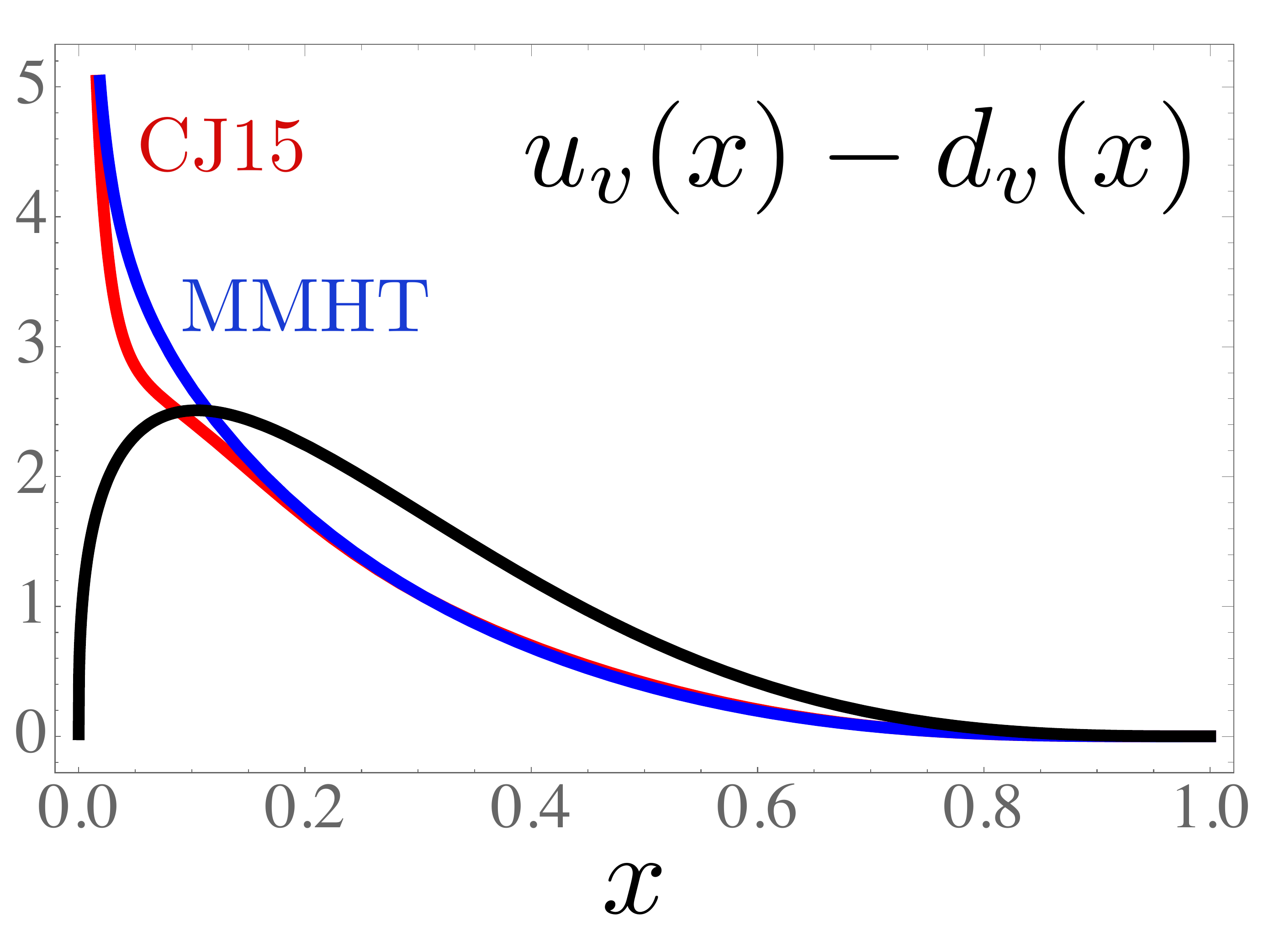}}
    \caption{Curve for $u_v(x)-d_v(x)$  at  $\mu=2.15$ GeV  built from the  data shown in Fig. \ref{Msbar16} 
 and   compared to CJ15 and MMHT global fits.
    \label{fabx}}
    \end{figure}

The reason for the discrepancy may be understood from  
Fig. \ref{fabx}, 
where 
 we compare 
 the normalized \mbox{$N x^{0.35} (1-x)^3\equiv q_v (x, \mu=2.15$ GeV)}  
 distribution to CJ15  \cite{Accardi:2016qay} and MMHT 2014 
   \cite{Harland-Lang:2014zoa} global fit PDFs, 
 taken at the  scale $\mu=2.15$ GeV. 
 Unlike the $\sim  x^{0.35} $ function, these PDFs are singular for small $x$,  
 which leads to the enhancement of  ITDs for large and moderate values of $\nu$.  

To fit the points for  ${\cal I}_R  (\nu, \mu^2) $, we have used the same 
simplest $N x^a (1-x)^b$ Ansatz for the PDF as in Ref. \cite{Orginos:2017kos}.   
In principle, one  may use 
more complicated models for PDFs  and get  practically the same fitted curve for the ITD   
in the $\nu \leq 4$ region, 
while a somewhat different curve for PDF $q_v (x)$.  
The reason is simple: the  inverse cosine Fourier transform is unique only 
when one exactly knows the ITD in the whole $0\leq \nu <\infty$ region.   
Performing such a transform from a limited $\nu \leq 4$ region,  
one needs to add some assumptions  either 
about the behavior of the ITD outside this region or about a functional form of the PDF $q_v (x)$. 
We fixed our choice by taking $q_v(x)\sim x^a (1-x)^b$. The   study of how  the 
shape of $q_v (x)$  varies if one uses more complicated forms, in particular, those used 
in the global fits  \cite{Accardi:2016qay,Harland-Lang:2014zoa}  is an interesting problem  
that, however,   goes  beyond the scope of the present paper.  

     Comparing to the LLA results 
     of Ref. \cite{Orginos:2017kos}, we observe 
     that   the large negative one-loop correction  in  Eq. (\ref{MNL})   has  visibly changed 
       the extracted PDF, which is  now    further 
       from the global fit  PDFs. 
       The main reason is that the $z_0=2a$ pseudo-ITD constructed in 
       Ref. \cite{Orginos:2017kos} 
       was treated there as corresponding to the $\mu\approx 1$ GeV  scale,
       while according to the modified rescaling relation (\ref{IMevL}),
       it   should correspond to $\mu \approx 4$ GeV.
       Hence, to get the $\mu \approx 2$ GeV curve,
       one needs to  evolve it  down in $\mu$.
       
       Still, the guiding idea of Ref. \cite{Orginos:2017kos},  that the 
       $\overline{\rm MS}$ ITDs ${\cal I}_R(\nu,\mu^2)$ can  be obtained from the 
       reduced pseudo-ITDs 
       ${\mathfrak R} (\nu,z_3^2)$ 
       by an appropriate rescaling $\mu = k/z_3$,  works with a rather good accuracy
   for all $z_3 \leq 6a$     if one takes $k\approx 4$.
    By this rescaling relation,    the $\mu  =1/2a  \approx 1$ GeV ITD  corresponds 
       to the $z_3 \approx  8a$ reduced pseudo-ITD.  As we discussed, 
        a boundary point beyond which the evolution stops, is $z_3 \approx 6a$.  
     Hence, the  pseudo-ITD   at  this distance  is   given by the 
       ITD ${\mathfrak R}_f (\nu)$  corresponding to the 
    universal fit  function $f(x)$ of \mbox{Eq.  (\ref{qV}).  } 
  This result  may be   also 
    obtained  by a direct  numerical calculation based on  \mbox{Eq. (\ref{MNL}). }
    
     Using Eq. (\ref{MNL}) one may also  evolve the 
    $\overline{\rm MS}$   ITD below $\mu=1/2a$, and the  resulting functions  will be
    changing with $\mu$.  On the other hand, the pseudo-ITDs do not change
    with $z_3$ when $z_3 \gtrsim 6a$. Hence,  the 
    rescaling connection ${\cal I}_R (\nu, \mu^2) \approx {\mathfrak R} (\nu, (4/\mu)^2)$
    in this region becomes less and less accurate when $\mu$ decreases,
    and eventually makes no sense.

\subsection{Imaginary part}

\begin{figure}[b]
  \centerline{\includegraphics[width=3.4in]{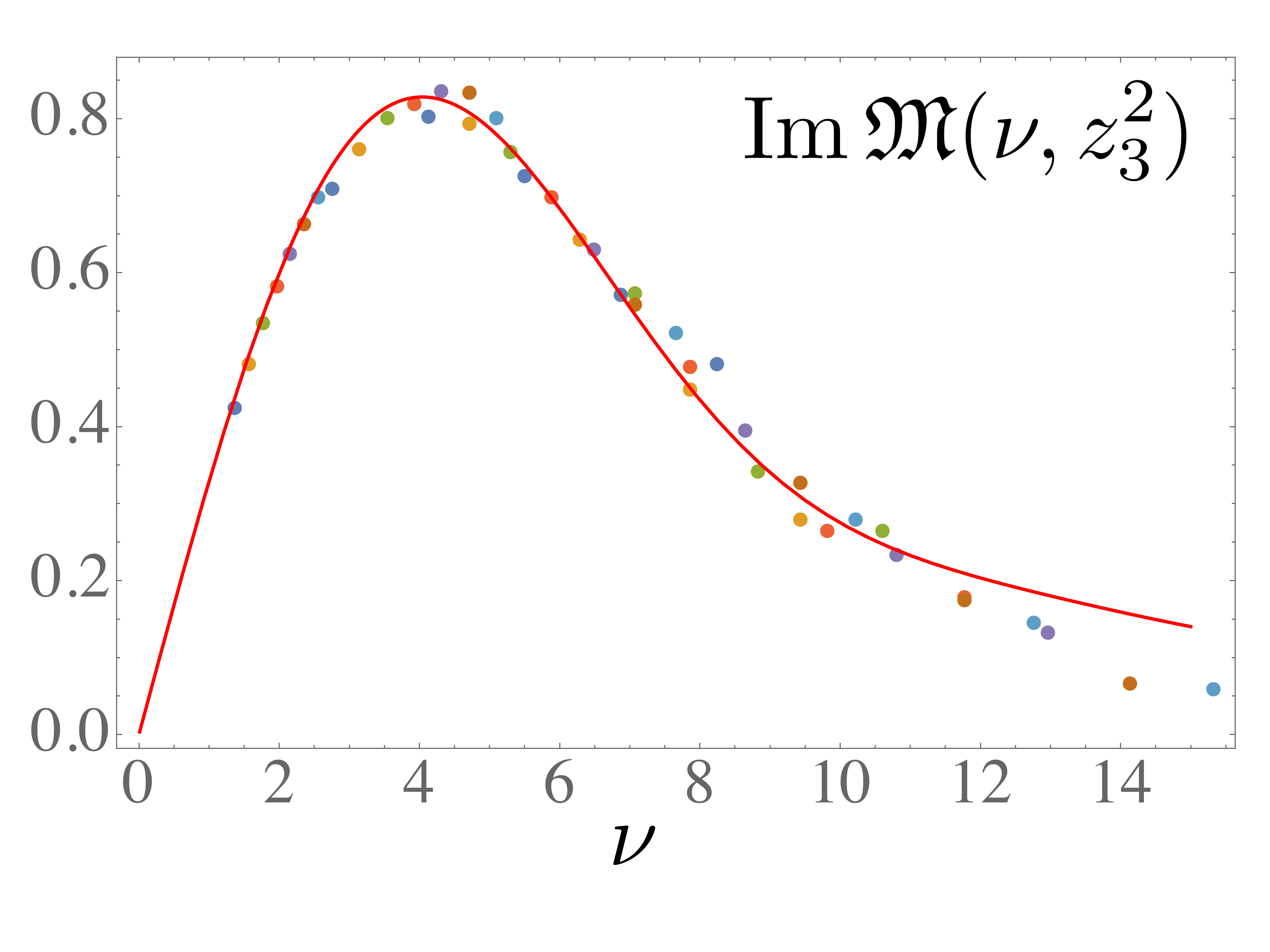}}
    \caption{Imaginary  part   of ${\mathfrak  M} (\nu, z_3^2)$  for  $z_3$  ranging  from  $7a$ to  $13a$.
    The curve corresponds to $q (x)  +  \bar q (x)= f(x) + 2 \bar q (x)$, with $f(x)$ given by Eq. (\ref{qV}) 
    and $\bar q (x)$ given by Eq. (\ref{barq}) 
    \label{imor713}}
    \end{figure} 

Imaginary part of the pseudo-ITD may be considered in a similar way.
 It  corresponds  to 
  the sine Fourier transform 
  \begin{align} 
{\rm Im} \,  {\mathfrak  M} (\nu)=  & \int_0^1 dx \,  
\sin  (\nu x)  \, [q (x)  +  \bar q (x)]
\label{MS}
     \end{align}
of the function   given by the sum   $q(x) + \bar q (x) $    of 
quark and antiquark   distributions.  This function 
differs from   the valence combination  $q_v(x) = q(x) -\bar q(x)$   by $2 \bar q (x) = 2[ \bar u (x) - \bar d(x)]$.
In Fig. \ref{imor713}, we show the data for large $z_3$ values $z_3 \geq 7a$. 
Just like in the case of the real part (see Fig. \ref{M713}), the points with $\nu \lesssim 10$
are close to a universal curve. Representing  $q (x)  +  \bar q (x) =q_v(x) +2 \bar q (x)$  
and taking $f(x)$ of Eq. (\ref{qV})   as $q_v (x)$,   we find  
  \begin{align} 
    \bar q (x) \approx  0.1 \,[20  x \, (1-x)^{3}]\,  . 
    \label{barq}
    \end{align}
Note that in  Ref.  \cite{Orginos:2017kos}, the fit was made 
for all the $z_3$ points (i.e. the points with $z_3 \leq 6a$ have been also included), and  the overall coefficient
for      $\bar q (x) $  was obtained to be 0.07 rather than 0.1.

\begin{figure}[t]
  \centerline{\includegraphics[width=3.5in]{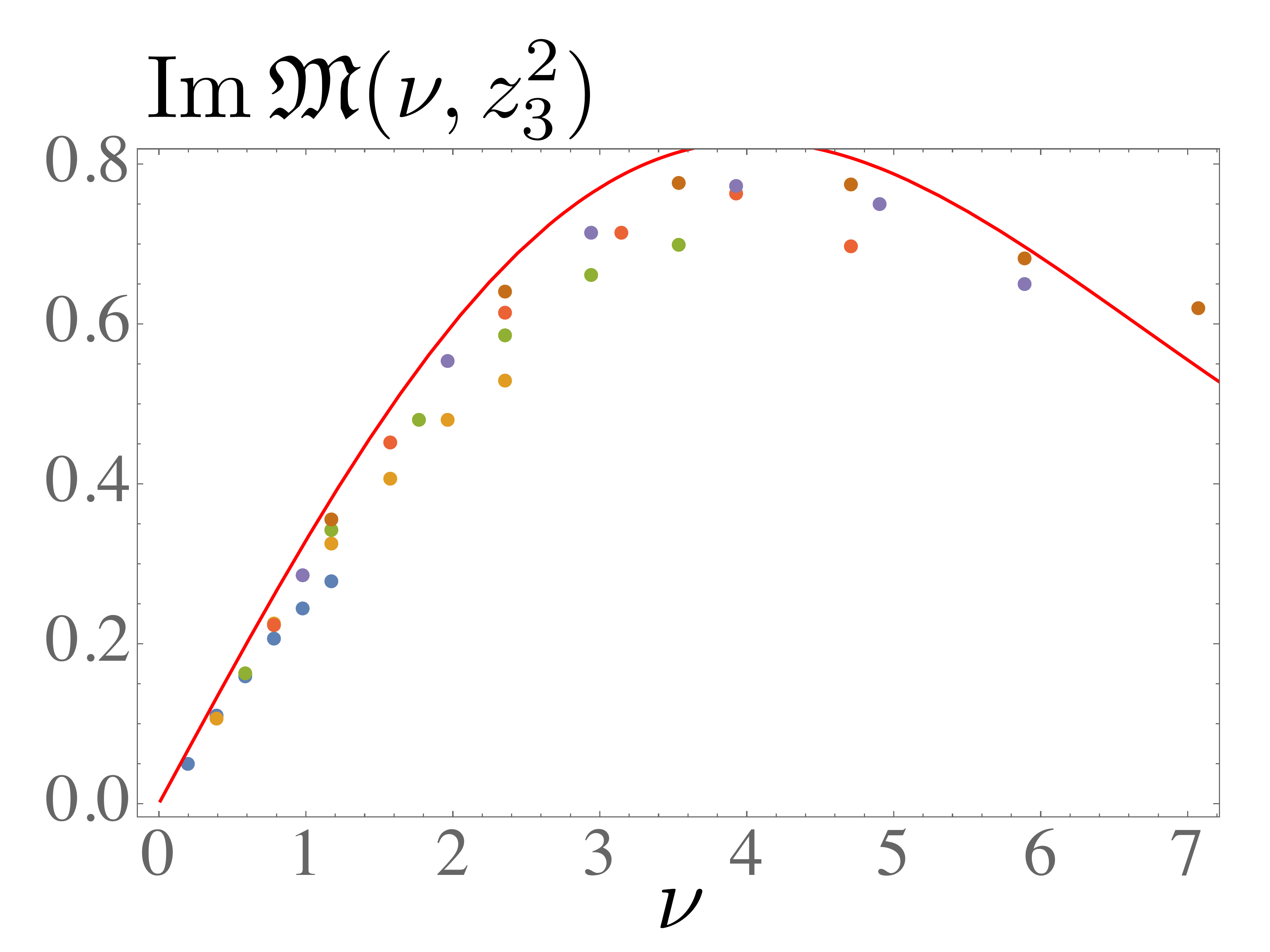}}
    \caption{Imaginary part   of ${\mathfrak  M} (\nu, z_3^2)$  for  $z_3$  ranging  from  $a$ to  $6a$.
The curve is the same as in Fig. \ref{imor713}. 
    \label{imor06}}
    \end{figure}

In Fig. \ref{imor06}, we show data with $z_3 \leq 4a$. As one can see,
all these  points are below the curve  obtained by fitting the $z_3 \geq 7a$ data. 
This is in agreement with the fact 
that, in the region $\nu \lesssim 6$, the perturbative evolution  
decreases   the imaginary part of the pseudo-ITD
when $z_3$ decreases. 
Note that the 1-loop  relation holds for the whole function ${\mathfrak M}= {\rm Re } \, {\mathfrak M} + i \, {\rm Im } \, {\mathfrak M}$.
So, we should just separate there real and imaginary parts, and the construction of 
the $\overline {\rm MS}$  function ${\rm Im}\,  {\cal I} (\nu, \mu^2) \equiv {\cal I}_I (\nu, \mu^2)$
proceeds in the same way  as for the real part. 

\begin{figure}[t]
  \centerline{\includegraphics[width=3.4in]{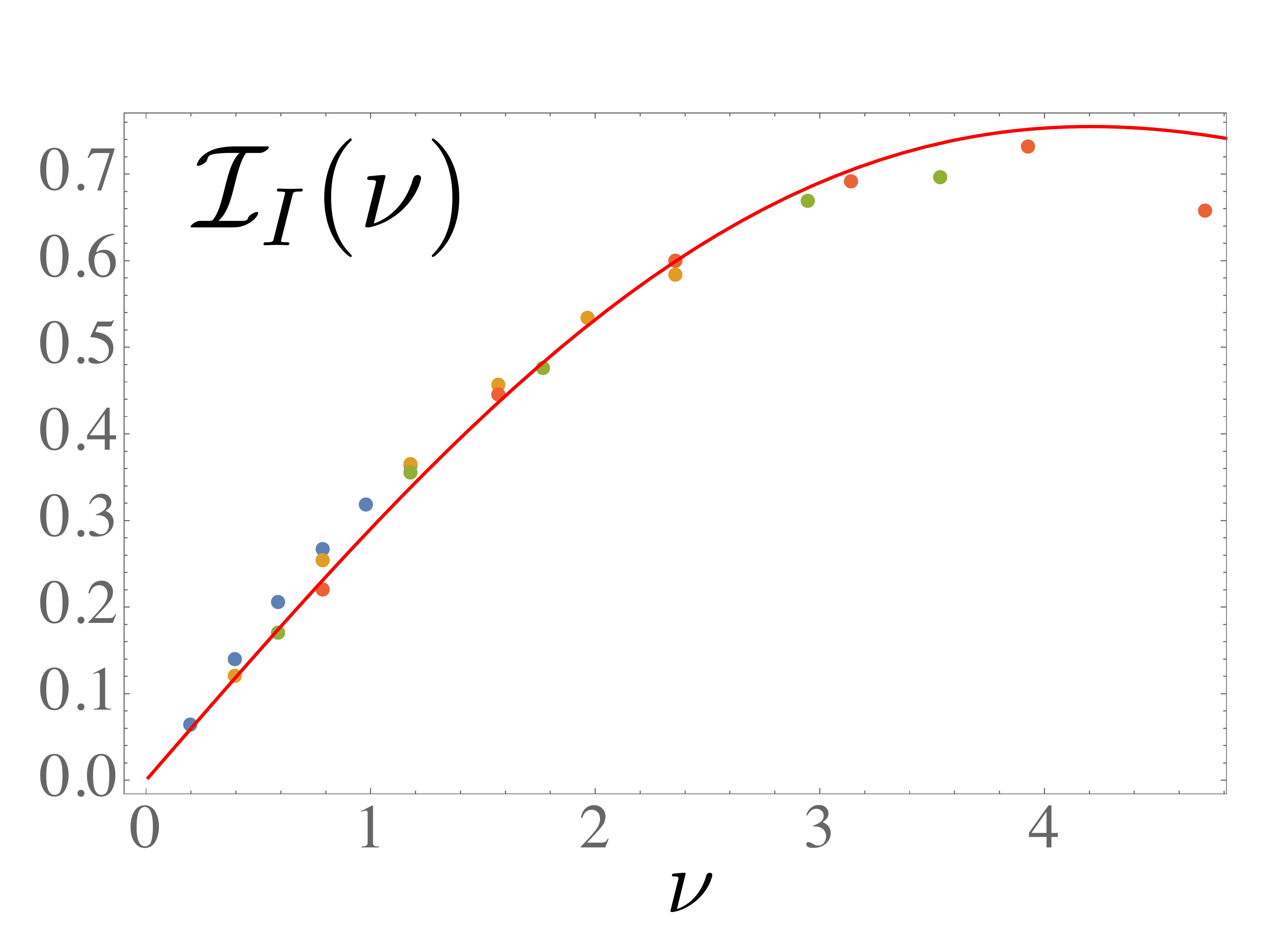}}
    \caption{Function  ${\cal I}_I  (\nu, \mu^2) $  for
    $\mu = 1/a$  calculated using  the data with   $z_3$   from  $a$ to  $4a$.
    The curve is described in the text. 
    \label{imev}}
    \end{figure}

 The results are shown in \mbox{Fig. \ref{imev}.}
Again, all the  points are rather close to a   universal    curve
   with a rather small scatter.  The curve shown corresponds to the sine Fourier transform 
   of the sum of the  valence distribution \mbox{$q_v (x, \mu =1/a) =N x^{0.35} (1-x)^3$} obtained 
  from the study of the 
   real part,  and the antiquark contribution $2 \bar q (x, \mu =1/a$). The latter was found from the fit  to be 
   given by $ \bar q (x, \mu =1/a=2.15$ GeV) $= 0.07 [20 x (1-x)^3]$.

   \section{Summary and conclusions}  
   
   In this paper, we  have extended   the leading-logarithm 
   analysis of lattice data for  parton pseudo-distributions 
   and reduced pseudo-ITDs 
   performed in Ref. \cite{Orginos:2017kos}. 
   To this end, we incorporated 
    recent  results for the reduced pseudo-ITDs at   the one-loop level   \cite{Radyushkin:2017lvu} 
   (see also \cite{Ji:2017rah,Izubuchi:2018srq}).

    It was  found that the correction contains a  large term
    resulting  in essential numerical  changes  compared to  the LLA.
    The   large correction appears since  
    effective distances involved in the most important diagrams
    are much smaller than the nominal distance $z_3$. 
    This  leads to a  change
    (from $k_{\rm LLA} \approx 1$ to $k \approx 4$  in the case of our
    particular ITDs)  of the 
    coefficient $k$ in  the 
    rescaling relation $\mu = k/z_3$  that allows to (approximately) 
    convert  the pseudo-PDFs ${\cal P} (x,z_3^2)$ into the $\overline{\rm MS}$
    PDFs $f(x,\mu^2)$.

 While the  rescaling relation serves as an  
instructive guide for quick estimates and  semi-quantitative analysis,
 the $\overline{\rm MS}$ ITDs may be directly constructed applying   the 
 exact one-loop formula.  Using it, we have obtained 
  the 
    ITD ${\cal I} (\nu, \mu^2)$ at the \mbox{$\mu=1/a \approx 2.15$ GeV}  
    $\overline{\rm MS}$ scale
          using the data in the  \mbox{$0\leq z_3 \leq 4a$}   region. 
          
         We  found that ${\cal I} (\nu, \mu^2)$ at this scale is  close
          to the reduced pseudo-ITD ${\mathfrak M} (\nu,z_3^2)$  for $z_3\sim 4a$.
          Since  all the data in the $a\leq z_3 \leq 4a$ region do not differ much from 
          the $z_3=4a$ ones (see Fig. \ref{M16}), the conversion of 
          the ${\mathfrak M} (\nu,z_3^2)$ data into ${\cal I} (\nu, 1/a^2)$
          does not involve large changes, i.e., 
  the perturbative expansion for   $\overline{\rm MS}$  ITD ${\cal I} (\nu, \mu^2)$
  in terms of the reduced 
    pseudo-ITDs ${\mathfrak M} (\nu,z_3^2)$    is  under control.  
 A  formal reason is that the   large  correction in this  case can  be 
    absorbed into the $z_3^2$-dependent evolution term, 
    with remaining corrections being small.  
         
       Phenomenologically,  
    the  PDF extracted in this way  follows the trend of  those
    given by the   global fits in the $x>0.1$ region,  but 
    does not reproduce their  singular behavior in the $x<0.1$ region.  
    The latter is usually related to the $x^{-0.5}$ pattern of the $\rho$-meson 
    Regge trajectory. Since the $\rho$-meson is essentially a rather narrow 
    resonance in the $\pi \pi$ system,  
    one should not expect to accurately reproduce the $\rho$-meson  
    properties  in a lattice simulation in which the pions 
    are as heavy as 600 MeV.  Thus,  one may hope  that using simulations 
    at physical pion mass would produce a better agreement with the global fits in the  
    small-$x$ region.  This hope is supported by recent extractions  \cite{Alexandrou:2018pbm,Chen:2018xof}  
    of $q_v (x)$ using  the quasi-PDF lattice simulations at physical pion mass. 

\acknowledgements

I  thank J. Karpie, K. Orginos and S. Zafeiropoulos,
my collaborators on  Ref. \cite{Orginos:2017kos}, 
 who performed the 
 lattice simulations,   the  results of which  were  analyzed in that paper 
 and also  used    in the present work. 
I am  especially  grateful to 
K. Orginos for  collaboration on  the pseudo-PDF evolution
in Ref.  \cite{Orginos:2017kos} and further discussions of this subject.
I thank  N. Sato 
for  providing the code generating 
the global fit PDFs  and Y. Zhao for discussions of one-loop corrections. 
This work is supported by Jefferson Science Associates,
 LLC under  U.S. DOE Contract \#DE-AC05-06OR23177 
and  by U.S. DOE Grant   \mbox{\#DE-FG02-97ER41028. }

\end{document}